\documentclass[aps,twocolumn,notitlepage]{revtex4-1}
\usepackage{amsfonts}
\usepackage{amsmath}
\usepackage{amssymb}
\usepackage{graphicx}
\providecommand{\U}[1]{\protect\rule{.1in}{.1in}}
\begin{document}
\title{Stripe Domains and First Order Phase Transition in the Vortex Matter of  Anisotropic High-temperature Superconductors}
\author{V. K. Vlasko-Vlasov$^{ 1}$, J. R.  Clem$^{ 2}$, A. E. Koshelev$^{ 1}$, U. Welp$^{ 1}$, W. K. Kwok$^{ 1}$}
\affiliation{1 Materials Science Division, Argonne National Laboratory, Argonne, Illinois 60439}
\affiliation{2 Department of Physics and Astronomy, Iowa State University, Ames, IW 50011-3160,USA}

\begin{abstract}

We report the direct imaging of a novel modulated flux striped domain phase in a nearly twin-free YBCO crystal. These domains arise from instabilities in the vortex structure within a narrow region of tilted magnetic fields at small angles from the in-plane direction.  By comparing the experimental and theoretically derived vortex phase diagrams we infer that the stripe domains emerge from a first order phase transition of the vortex structure.  The size of domains containing vortices of certain orientations is controlled by the balance between the vortex stray field energy and the positive energy of the domain boundaries.  Our results confirm the existence of the kinked vortex chain phase in an anisotropic high temperature superconductor and reveal a sharp transition in the state of this phase resulting in regular vortex domains.  

\end{abstract}
\maketitle

In type II superconductors (SC) the magnetic field penetrates in the form of vortices - magnetic flux tubes each carrying a single flux quantum \cite{1}.  Interactions between vortices and their coupling to crystal defects result in a rich variety of vortex structures.  They include a triangular or square vortex lattice that can subsequently melt at high temperatures to a vortex liquid, a pinned vortex glassy state that can sustain high current carrying capacity, and in highly anisotropic superconductors, 2D pancake vortices that can interact with in-plane Josephson vortices to create 1D vortex chain states.  These states of \textit{vortex   matter} define all transport and magnetic properties of applied superconductors.  

Here we report on an unusual vortex domain structure that arises following a first order phase transition in the vortex state in nearly untwined YBCO crystal under tilted magnetic fields. Typically, vortices repulse each other when the distance between them is much smaller than the penetration depth $\lambda$, the scale over which the magnetic field of the vortices decays.  In this case, the vortices can arrange into a uniformly spaced lattice or form smooth density gradients defined by the critical current, the maximum current that the superconductor can sustain before reverting to the normal state.  In contrast, at larger distances the interactions between vortices can be \textit{attractive}, resulting in the formation of vortex chains and bundles  (see review \cite{2}). For example, periodic stripe domains and circular bundles of vortices interspersed with flux free \textit{Meissner} regions occur in thin niobium disks with a low Ginzburg-Landau parameter $\kappa =\lambda/\xi\sim1$, where $\xi$ is the coherence length  setting the size of the vortex core.  These vortex domains are similar to the alternating normal (N) and superconducting Meissner (M) domains found in the intermediate state of type I SCs with a nonzero demagnetizing factor.  In the latter, the lamellae M/N domains are known to minimize the magnetostatic energy of the normal regions at the expense of the \textit{positive} energy of the boundaries between the normal and SC phases. 

Vortex attraction can also appear in large $\kappa$, anisotropic type II SCs when an applied magnetic field is tilted from the anisotropy axis \cite{3,4,5}.  Here, the circulating supercurrents that form the vortex tend to flow perpendicular to the anisotropy axis rather than to the vortex line.  This creates an inversion of the magnetic field at some distance from the vortex core resulting in an attraction of the neighboring vortex within the tilt plane \cite{3, 5} and formation of dense vortex chains coexisting with the dilute lattice of Abrikosov vortices, as observed in decoration, Hall probe, and electron microscopy experiments in layered high-Tc superconductors (see review \cite{6}).  

Formation of patterns including stripe structures is general for a wide variety of condensed matter and colloidal systems with competing attractive and repulsive interactions, which are usually extended at different length scales \cite{7,8,9}. In fact, long range attraction and short range repulsion of vortices in superconductors with two order parameters [10] as also in multilayers of different superconductors \cite{11,12} can result in formation of vortex stripes. Experimentally, stripes of vortices intermittent with Meissner phase were observed in $MgB_{2}$ single crystals \cite{13,14} and were treated as a possible consequence of the two different SC condensates in the 2-band magnesium diborate. 

In this work we discovered a new vortex structure of regular stipe domains with alternating flux density, $B_{z}$, generated by magnetic fields tilted from the ab-plane in a nearly twin-free YBCO platelet crystal.  Unlike vortex/Meissner domains and vortex chain structures, our vortex domains are created via an instability in the vortex state \cite{15,16,17,18,19,20,21,22} due to a first order phase transition characterized by an abrupt jump in the vortex tilt angle \cite{23}. A possibility of such first order transition in the vortex orientation was first noticed by Buzdin and Simonov in \cite{24}. During the phase transition, an intermediate state with domains of different vortex phases (orientations) emerges.  We propose that similar to the case of type I SCs and ferromagnets, the regular structure of domains with alternating density of $B_{z}$ reduces the magnetostatic energy of the vortex stray fields perpendicular to the surface of the sample. Between the domains with different flux orientations vortices gradually rotate forming a domain boundary. Responsible for this rotation supercurrents yield a positive domain wall energy which limits the refinement of domains caused by magnetostatics. We suggest that the attractive coupling of tilted vortices into chains along the vortex tilt plane \cite{3,4,5} and attraction between the neighboring chains with mutually shifted vortices stabilizes domains with larger $B_{z}$.  Our calculated vortex structure phase diagram in tilted fields defines the boundaries of the angular vortex instability in YBCO in qualitative agreement with the experiment.

The  $YBa_{2}Cu_{3}O_{7-\delta}$ rectangular platelet crystal (1130x340x20 $\mu m^{3}$) used in our experiments was grown using a flux method.   The crystal had a few narrow twin lamellae near one end but most of the area was twin-free.  The onset temperature of the superconducting transition and the width were $T_{c}$ = 92.4K and $\Delta T$ = 0.3K respectively, as determined from magnetization measurements.  We imaged the perpendicular magnetic fields ($B_{z}$) on the sample surface using a magneto-optical (MO) indicator technique \cite{25}.  Data were obtained after cooling the sample in applied magnetic fields tilted from the ab-plane of the crystal as well as during remagnetization of the sample in tilted fields.  The resulting intensity of the MO images depicts the strength and polarity of the normal field $B_{z}$, 
 Fig.1a delineates the spatial field pattern at T = 60K after cooling the sample in a field of $H_{//}$ = 1388 Oe parallel to the long edge of the sample and then switching off the field.  The dark (negative $B_{z}$) and bright (positive $B_{z}$) image contrast on the right and left edges of the sample demonstrates that trapped vortices induced by the applied field lie parallel to the sample plane and their ends produce stray fields that diverge up at the left edge and converge down at the right edge as shown in Fig.1d. Similar images were observed when $H_{//}$ was tilted by less than  $\sim 0.5^{\circ}$ from the sample plane. 

At tilt angles $> 0.5^{\circ}$ a qualitatively new stripe pattern emerges at the center of the sample. Fig.1b shows such a pattern formed after cooling the sample in a field $H_{//}$ = 1388 Oe with a tilt angle of  $\sim 1.60^{\circ}$ and switching off the field. Similar to Fig.1a there is bright and dark contrast at the short sample ends and a new dark contrast  appears at the top and bottom long edges of the crystal.  The latter is associated with negative stray field around the sample due to the positive $B_{z}$ trapped in the central region of the sample. It is in this region the periodic stripes with alternating $B_{z}$  values reside.  Fig.1e shows the $B_{z}$  profile measured across the stripes. It clearly reveals periodic oscillations of Bz superimposed on a monotonic spatial induction gradient. The latter is typical of perpendicular trapped flux in a superconducting plate and corresponds to the average critical current circulating in the ab-plane of the sample \cite{26,27,28}. We associate the observed stripe pattern with a domain structure of vortices with different tilt angles or, equivalently, with different density of in-plane and $\it{c}$-axis vortex segments within the vortex staircase structure inherent for cuprates \cite{29, 30, 19, 23} (see Fig.1f). 	This novel domain structures exist in a narrow range of tilt angles between $\theta \sim 0.5^{\circ}$ and $2.40^{\circ}$. Within this range, the width and length of the trapped $B_{z}$ region and thus the length of the stripe domains were slightly shorter for smaller angles.  At a fixed angle the trapped $B_{z}$ region was shorter for smaller fields. Fig. 2 depicts the phase diagram of the observed stripe domain structures obtained by cooling the sample to 60K under various angles of applied fields $H_{//}$ ranging from $\sim$280 to 1388 Oe. 
\begin{figure*}[htb]
\vspace{0.5cm}
\includegraphics[width=10.0cm]{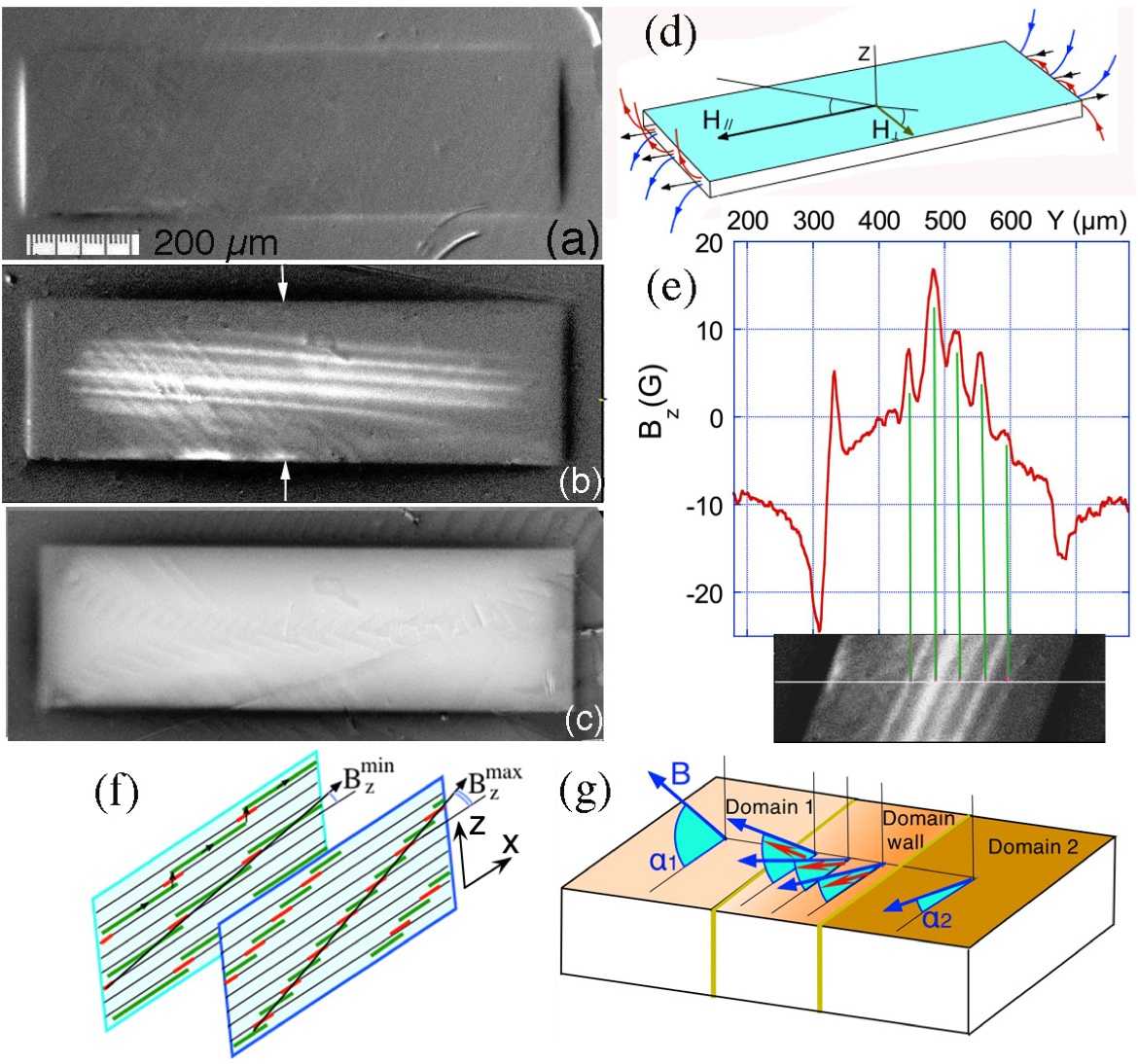}
%\hspace{-5cm}
\vspace{-0.5cm}
\caption{(a) Trapped flux (H=0) after field cooling from $T > T_{c}$ to 60K in $H_{//}$  = 1388 Oe. (b) Same after cooling in H = 1388 Oe tilted by $1.6^{\circ}$. (c) Same after cooling in H = 1388 Oe tilted by $3.5^{\circ}$. (d) Profile of $B_{z}$ between two arrows shown in Fig.1b. (d) Scheme of the stray fields corresponding to (a). (e) $B_{z}$ profile across the sample between arrows shown in (b). (f) Scheme of the staircase vortices in cuprates. (g) Scheme of the vortex domains. Currents (short arrows) in the domain wall flow along vortices in force free configuration. }
%\label{fig:Fig.A}
\end{figure*}

To track the initial stage of the domain nucleation we imaged the flux structure during gradual reduction of the field after initially field cooling the sample to T = 60K with $H_{//}$ = 1388 Oe tilted by $1.6^{\circ}$ (see MO pictures in Supplemental Info).  Before reducing $H_{//}$, the $B_{z}$ distribution  is homogeneous over the entire imaged area.  Decreasing field to $\sim$ 900 Oe results in the appearance of a dark edge contrast due to the stray fields coming from the trapped \textit{positive} $B_{z}$. Simultaneously, stripes of alternating bright and dark contrast (modulated $B_{z}$) emerge along $H_{//}$ near short ends of the sample.  Further decreasing the magnetic field, extends the stripes over the entire length of the sample. At even smaller $H_{//} \sim$150 Oe,  vortices turn into the plane near the perimeter of the sample and a region of trapped $B_{z}$ with stripe domains forms inside the crystal. This configuration remains after reducing the field to zero.  For angles larger than $2.4^{\circ}$ the field cooling resulted in a larger region of stronger trapped $B_{z}$ but without any stripe domain features (Fig.1c). 
\begin{figure}[htb]
\vspace{0.5cm}
\includegraphics[width=8.5cm]{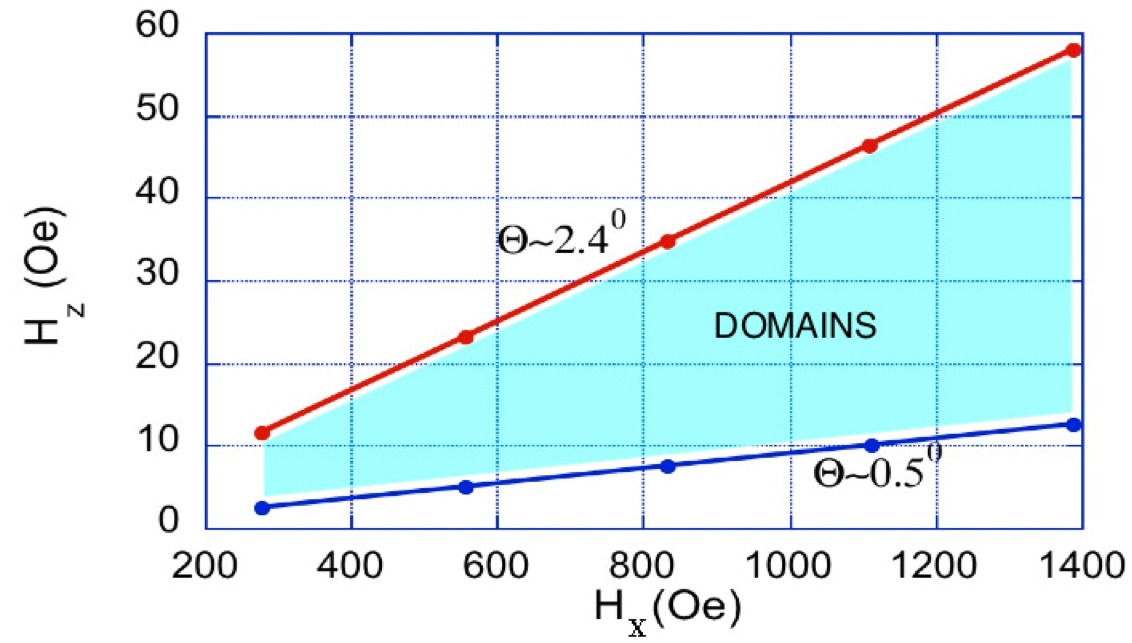}
%\hspace{-5cm}
\vspace{-0.5cm}
\caption{Experimental $H_{z}-H_{x}$ domain existence map. }
%\label{fig:Fig.A}
\end{figure}

The nucleation of domains could be also initiated if after cooling in $H_{//}$=1388 Oe the field was first increased by a few hundred gauss and then decreased back to the initial value. After field cooling the MO picture was homogeneous. But after increasing and decreasing field back to 1388 Oe, stripes appeared at the short ends of the sample. They extended over the central area during further reduction of $H_{//}$ as described above.

Similar observations were also performed in tilted fields $H_{\perp} (\theta_{\perp})$ perpendicular to the long sample side.  In this case the stripe domains formed along the width of the sample and were wider and somewhat less regular than in $H_{//}$ (Fig.4a). They also exist in a narrow range of tilt angles ($\theta_{\perp}\sim$ 0.4 to $3^{\circ}$) and become shorter with decreasing $\theta_{\perp}$ (Fig.4b). At smaller angles the total trapped $B_{z}$ is weaker and black and white contrast at the long edges of the crystal appears due to the stronger in-plane flux $B_{y}$.  The domains emerge in the shape of narrow wedges stretching along $H_{\perp}$ from the long edges during application of large enough field. At decreasing field they detach from the edges and remain in the central region after switching $H_{\perp}$ off.

So far there were no experimental reports of vortex domains in high-$\kappa$ SCs. The coexistence of vortex lattices with different tilt angles could be expected from theoretical predictions of vortex angle instabilities. These instabilities were calculated for some range of angles for (i) a single tilted vortex line in anisotropic three-dimensional superconductors \cite{15,16,17, 20,21,22}, (ii) a single tilted vortex line in layered superconductor with purely magnetic coupling\cite{19}. and (iii) a chain of tilted vortex lines \cite{23}. Below, we derive the vortex phase diagram for YBCO under tilted magnetic fields by taking into account the energy of individual vortices and energies of their interactions and their coupling to external fields.  We then analyze the stray fields in thin SC plates and show that they define the width of the stripe domains and discuss vortex attraction as a possible mechanism for stabilizing domains with larger $B_{z}$.

\begin{figure}[htb]
\vspace{0.5cm}
\includegraphics[width=7cm]{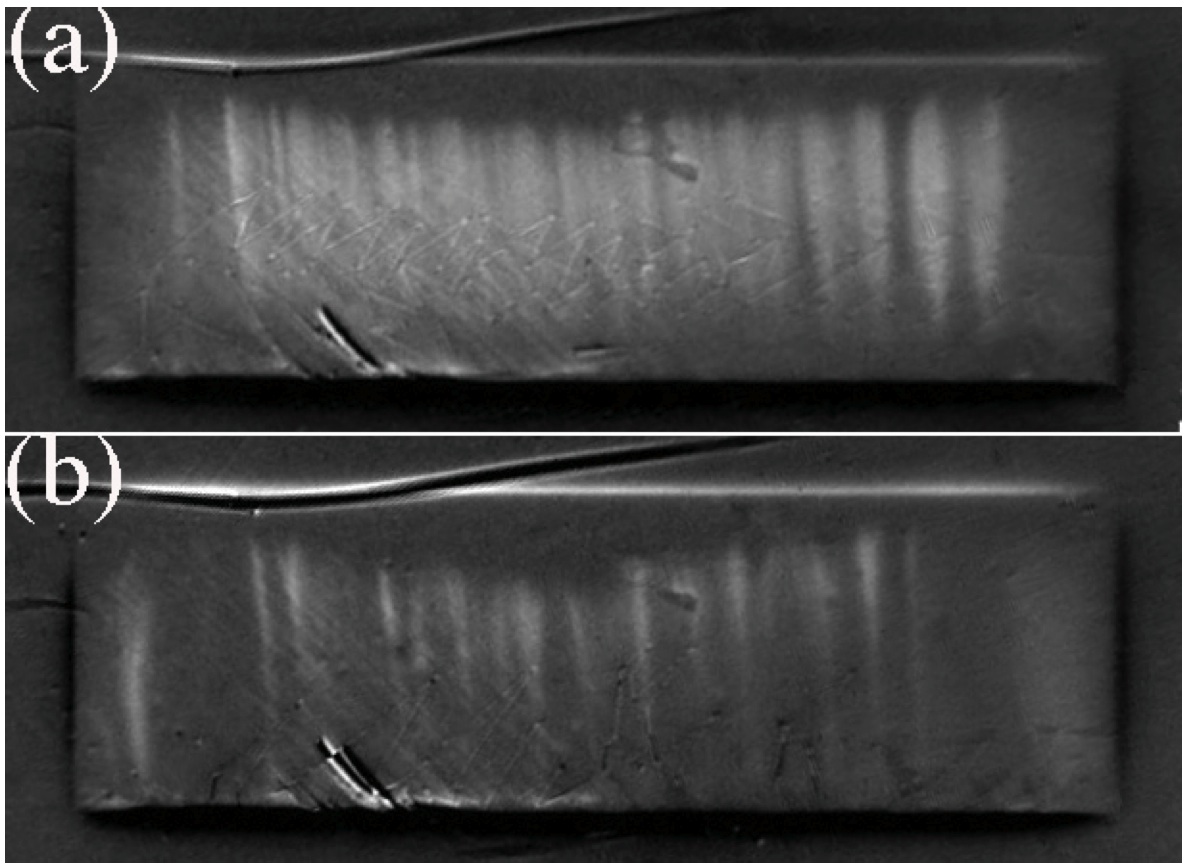}
%\hspace{-5cm}
\vspace{0.5cm}
\caption{Vortex domains at 60K after application and switching off $H_{\perp}$=2100 Oe along the short side tilted from the ab-plane by $\sim  1.20^{\circ}$ (a) and $\sim 0.40^{\circ}$(b) }
%\label{fig:Fig.A}
\end{figure}
The vortex phase diagram in tilted fields is constructed using analytic results for vortex-chain energies derived in Ref. \cite{23} (see \cite{31} for details). Our calculations for a bulk sample show abrupt changes in the vortex angle which are mostly driven by the energy of an isolated vortex chain formed \textit{in   the    tilt    plane} while interactions between the chains may be neglected.  The total energy of a vortex chain consists of isolated vortex line and inter-vortex coupling contributions, $E_{TV} = E_{TV}^{s} + E_{TV}^{i}$.  Both terms depend on the in-plane and $\it{c}$-axis magnetic field components and on the SC materials parameters \cite{31}.  The structure of vortices tilted by a small angle from the in-plane direction, $\theta < 1/\gamma$, can be described by the addition of $\it{c}$-axis pancake vortices (kinks) to the Josephson vortex line lattice in purely in-plane field. The equilibrium state is obtained by minimizing the sum of $E_{TV}$ and the energy of interaction of the $\it{c}$-axis flux with thermodynamic magnetic field $H_{z}$. We found that in some range of parameters the total energy has minima at two values of $B_{z}$ and the system undergoes a first-order phase transition at the critical value of $H_{z}$ resulting in a jump in $B_{z}$. Roughly, the transition is located in the in-plane field range $B_{x} < \gamma \Phi_{0}^{2} /\lambda_{ab}^{2}$ and bypassed range of $B_{z}$ corresponds to tilt angles around $\theta \sim 1/\gamma$. Fig. 4 shows numerically computed phase diagram obtained for  typical YBCO parameters, the penetration depth $\lambda_{ab}$ =200 nm, distance between $CuO_{2}$ layers $\it{s}$= 1.2 nm, and anisotropy $\gamma$ = 7. In the region between two lines in Fig.4 the vortex structure experiences an angular instability and the c-axis component of the flux, $B_{z}$, jumps from $B_{z}^{min}$ to $B_{z}^{max}$ via a first order phase transition. Although the sample shape and vortex pinning are not accounted for in this theory, the calculated phase diagram is in reasonable agreement with  our experimental $H_{z}-H_{x}$ data of the emergence of the stripe domains (Fig.2). The experimental $H_{z}$ fields limiting the domain existence are somewhat smaller compared to $B_{z}$ of the theoretical instability lines. We attribute this difference to the demagnetization factor of the sample and infer the angular vortex instability accompanying the vortex angle transition as the main reason for the appearance of domains.
\begin{figure}[htb]
\vspace{0.5cm}
\includegraphics[width=8.5cm]{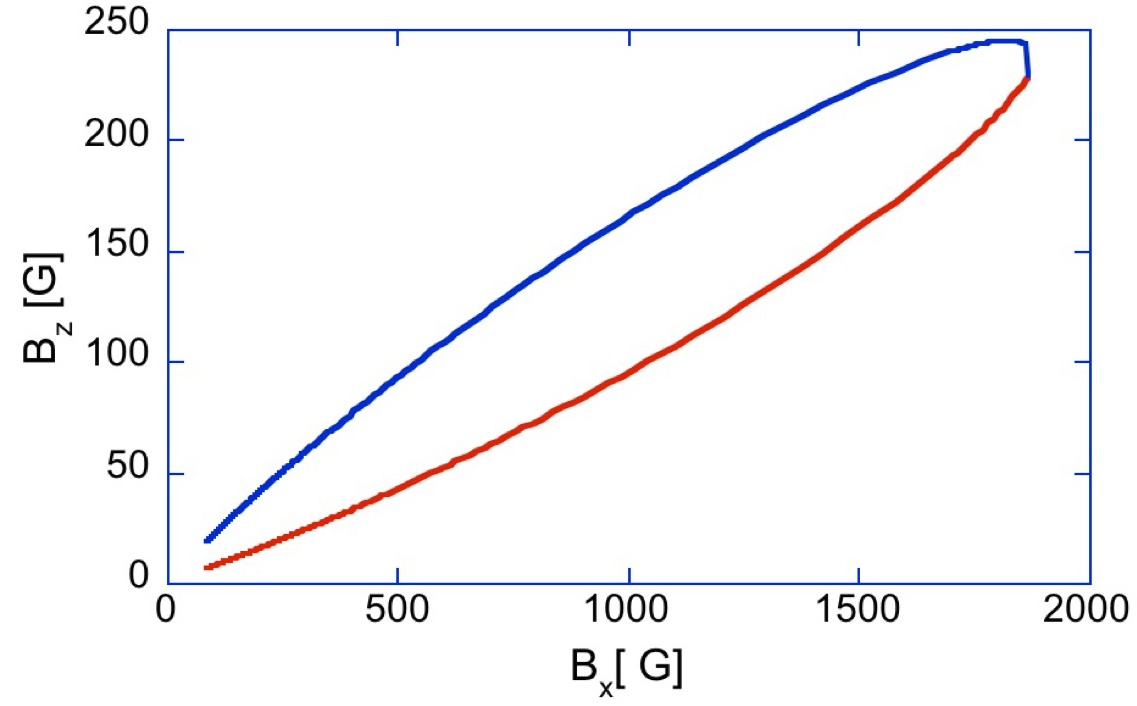}
%\hspace{-5cm}
\vspace{0.5cm}
\caption{Calculated phase diagram of the vortex state in YBCO under tilted fields. With increasing tilt, the $B_{z}$ component jumps from the bottom to top line. The intermediate state domains should form in the area between the lines. }
%\label{fig:Fig.A}
\end{figure}

In large SC samples the sum of energies of individual vortices and bulk vortex-vortex interactions is the main factor defining the flux state. However, in thin SC plates the external stray fields induced by vortices come into play.  First, they define the long-range vortex interactions near the surface, which can be treated as a repulsion of magnetic monopoles  \cite{32}. Second, they can introduce modulated flux structures, which would reduce the energy of the vortex stray fields around the superconductor ($E_{s}$). In type I SCs and low-$\kappa$ materials this energy is responsible for the emergence and size of the SC/N or Meissner/Shubnikov lamella domain structures  \cite{2}. We conjecture that the regularity of our vortex stripe domain structure can also be explained by the effect of the stray field energy, $E_{s}$. Similar to the case of ferromagnetic domains, $E_{s}$ can be minimized by periodic oscillations of $B_{z}$. In relatively small fields, we can consider our system as comprised of up and down magnetized domains superimposed onto an averaged trapped field background. The latter is not important for the variational problem.  Formulas for the stripe domains give $E_{s} \sim1.7\mu_{0}^{-1}(\Delta B_{z}/2)^{2}D$ \cite{33}. Here $\Delta B_{z}$ is the difference of $B_{z}$ in neighboring domains and $D$ is the domain width.  An important factor limiting the refinement of domains is the \textit{positive} energy at their boundaries $\sigma_{B}$. The balance between the reduction of the stray field energy and the increase of the boundary energy yields the equilibrium size of the domains. Positive $\sigma_{B}$ is produced by currents responsible for the rotation of vortices in the boundary layer between domains with different flux direction.  For example, let us consider two infinite blocks of vortices of the same density $B_{0}$ turned by an angle $\alpha_{0}$ ($\alpha_{1}- \alpha_{2}$ in Fig.1f) with a $x_{0}$-thick boundary between them where $B_{0}$ rotates in a force-free way ($J(x)//B(x)$), so that $B(x)= \hat{y} B_{0}sin(\alpha (x))+\hat{z} B_{0}cos(\alpha (x))$. The current $J(x)$ flows only within the boundary layer and vanishes in the homogeneous domains. In the boundary region the current density is $J(x)=\mu_{0}^{-1}B_{0}(d\alpha/dx)$.  For a linear variation of the vortex angle $\alpha(x) = x(\alpha_{0} /x_{0})$ , the force-free current $J(x)$ has some maximum density $J_{c//}$ with the kinetic energy density $\mu_{0}\lambda^{2}J_{c//}^{2}/2$.The resulting energy per unit area of the boundary will be  $\sigma_{B}=J_{c//}B_{0}\alpha_{0}\lambda^{2}/2  > 0$. Obviously, the real structure of the boundary is more complex, but the energy due to the boundary currents should still be positive. Minimizing the energy per unit area of the plate $E_{S} +E_{B} = 1.7\mu_{0}^{-1}(\Delta B_{z}/2)^{2}D+\sigma_{B} (d/D)$, where $d$ is the thickness of the sample, yields the equilibrium size of the stripe domains, $D = (4\mu_{0}d \sigma_{B}/ 1.7\Delta B_{z}^{2})^{1/2}$.  Remarkably, this simple formula gives the domain width $D=30 \mu m$ (for $d=20 \mu m, J_{c//}=10^{7} A/cm^{2}, \alpha_{0}=2^{\circ}, \lambda=200 nm$ and $\Delta B_{z}$=10 G)  very close to the observed  values ($D\sim 20 \mu m$).

One could expect that in the increased Bz domains there is an enhanced repulsion of vortices. However, we suggest that the vortex attraction in the chains along the tilt plane and attraction between the chains can stabilize these domains. For a linear chain of tilted vortices the intrachain attraction emerges through the pancake components of vortices and has a dipolar character \cite{34}. The same dipolar attraction can occur between the chains if pancake stacks in the neighboring chains are shifted by half-a-period along the chain. A possible confirmation of this scenario could be zigzag vortex chains observed in $NbSe_{2}$ in fields close to the basal plane \cite{35}. This mechanism can also account for the formation of chain bundles in BSCCO in fields strongly tilted from the c-axis \cite{36}.  

In summary, we report the first direct imaging of a novel vortex stripe domain phase in a nearly twin-free YBCO crystal.  The domains with alternating $B_{z}$ arise from instabilities in the vortex state within a narrow region of tilted magnetic fields at small angles from the ab-plane.  By comparing the experimental and theoretically derived vortex phase diagrams we infer that the stripe domains emerge due to a first order phase transition of the vortex structure. We show that the size of domains with different vortex angles is defined by the balance of the vortex stray field energy and positive energy of domain boundaries.  Our results confirm the emergence of the kinked vortex chain phase in an anisotropic high temperature superconductors and reveal sharp transformations in their state.

\vspace{1cm}
\textit{Acknowledgements} 
This work was supported by the US Department of Energy DOE BES under Contract No. DE-AC02-06CH11357. One of us (JR) acknowledges support from the Center for Emergent Superconductivity, an Energy Frontier Research Center funded by the US Department of Energy, Office of Science, Office of Basic Energy Sciences.

%\vspace{-5cm}

%\vspace{20 cm}
\newpage
\begin{center}
\appendix
%\maketitle{\bf\large{Phase diagram for tilted chains}}
\section{Phase diagram for tilted chains}
%\vspace{0.5cm}
\maketitle\small{Supplemental Info for "Stripe Domains and First Order Phase Transition in the Vortex Matter of  Anisotropic High-temperature Superconductors" by V. K. Vlasko-Vlasov$^{ 1}$, J. R.  Clem$^{ 2}$, A. E. Koshelev$^{ 1}$, U. Welp$^{ 1}$, W. K. Kwok$^{ 1}$}

\maketitle{\it{$^{ 1}$ Materials Science Division, Argonne National Laboratory, Argonne, Illinois 60439}}

\maketitle{\it{$^{ 2}$ Department of Physics and Astronomy, Iowa State University, Ames, IW 50011-3160,USA}}

\end{center}
\vspace{0.5cm}
Several theoretical studied demonstrated that in anisotropic superconductors
some orientations of dilute tilted vortex lattice are unstable [1s, 2s]. As a consequence, there is a first order phase
transition where the lattice orientation angle jumps. We analyze behavior of
the equilibrium c-axis induction $B_{z}$ for uniformly tilted vortex lattice
and find range of parameters where $B_{z}$ has a jump. We consider tilted
lattice composed of tilted chains separated by the distance $c_{y}$. The
in-plane distance between tilted vortices in a chain is $a$ and the $c$-axis
distance is $c_{z}$, see Fig. 5, so that the tilting angle with respect to the
$c$-axis is given by $\tan\theta=a/c_{z}$. The components of magnetic
induction are given by $B_{x}=\Phi_{0}/c_{z}c_{y}$ and $B_{z}=\Phi_{0}%
/ac_{y}=B_{x}c_{z}/a$.
\begin{figure}[htb]
%\vspace{-0.5cm}
%\includegraphics{FigA.jpg}
\includegraphics[width=4.0cm]{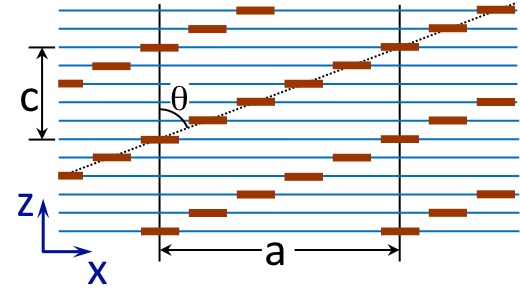}
%\hspace{-5cm}
\vspace{-0.5cm}
\caption{Scheme of tilted vortices.}
%\label{fig:Fig.A}
\end{figure}
To facilitate calculations, we make several simplifying assumptions. We
consider the range of small c-axis fields $B_{z}\sim B_{x}/\gamma$. This
allows us to neglect weak $B_{z}$ dependence of separation between the chains
and take the lattice parameters $c_{z}$ and $c_{y}$ to be the same as for the
Josephson vortex lattice for the field aligned with the layers. Two aligned
configurations of the Josephson vortex lattice are possible which are
degenerate within the London model. The $c$-axis lattice parameter is
connected with the in-plane magnetic field by relation $c_{z}=\sqrt{\beta
\Phi_{0}/\left(  \gamma B_{x}\right)  }$ with $\beta=2\sqrt{3}$ and
$2/\sqrt{3}$ for these two configurations. Analysis of commensurability
oscillations at high fields [3s] suggests that the first
configuration is preferable. Our analysis shows that discontinuous behavior
exists in the range $c_{z}\gtrsim\lambda_{ab}/\gamma$. In this range the
in-plane lattice parameter $c_{y}$ exceeds the London penetration depth and
interaction between the tilted chains gives only small contribution to the
total energy. The ground-state c-axis vortex density is mostly determined by
the energy of an isolated tilted chain and we focus on the analysis of this
energy [2s].
\begin{figure}[htb]
\vspace{0.5cm}
%\hspace{3cm}
%\includegraphics{FigA.jpg}
\includegraphics[width=8.0cm]{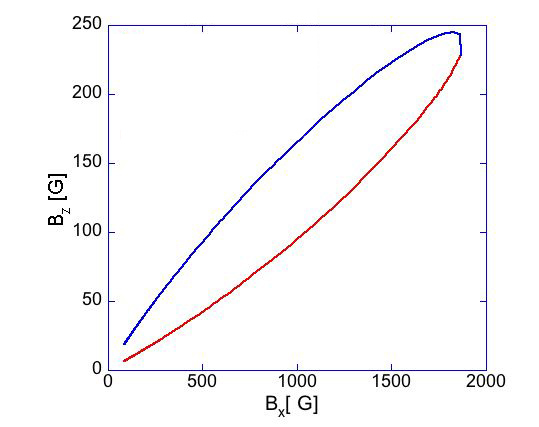}
\vspace{-0.5cm}
\caption{Phase diagram of the vortex state in YBCO in tilted fields. Between the curves vortices are unstable.   }
%\label{fig:Fig.A}
\end{figure}
The energy of an isolated tilted chain is composed of two parts, the energy of
isolated tilted vortices and the interaction term, $E_{TV}=E_{TV}^{s}%
+E_{TV}^{i}$. To facilitate numerical evaluation, we present the dimensionless
energy $\tilde{E}_{TV}=\lambda_{ab}E_{TV}/\varepsilon_{0}$ with $\varepsilon
_{0}=\Phi_{0}^{2}/(4\pi\lambda_{ab})^{2}$ as function of the reduced planar
density of pancake vortices $\tilde{n}=\lambda_{ab}/a$ and the lattice
parameter $\tilde{c}=c_{z}\gamma/\lambda_{ab}$. The field components are
related to these reduced parameters as%
\begin{equation}
B_{x}=\frac{\beta\gamma\Phi_{0}}{\lambda_{ab}^{2}\tilde{c}^{2}};\ B_{z}%
=\frac{\beta\Phi_{0}\tilde{n}}{\lambda_{ab}^{2}\tilde{c}}.\label{BxBz}%
\end{equation}
Analytical expression for the energy of an isolated tilted vortex is only
available for two asymptotic regimes [2s]:%
\begin{widetext}
\begin{equation}
\tilde{E}_{TV}^{s}\approx\tilde{E}_{PS}^{s}+%
%TCIMACRO{\QDATOPD{\{}{.}{\tilde{E}_{JV}^{s}+\tilde{n}\left(  \ln\frac{\gamma
%}{l_{s}}-0.81+\frac{\tilde{n}\tilde{c}}{2}\left[  \ln\left(  l_{s}\tilde
%{n}\tilde{c}\right)  -\frac{3}{2}\right]  \right)  \text{ for }\tan\theta
%\gg\gamma}{\tilde{n}\ln\frac{\sqrt{1+\left(  \gamma/\tilde{n}\tilde{c}\right)
%^{2}}+1}{2}+\frac{\ln\left(  8l_{s}\tilde{n}\tilde{c}\right)  -\gamma_{E}%
%}{2\tilde{n}\tilde{c}^{2}}\text{ for }\tan\theta\ll\gamma}}%
%BeginExpansion
\genfrac{\{}{.}{0pt}{0}{\tilde{E}_{JV}^{s}+\tilde{n}\left(  \ln\frac{\gamma
}{l_{s}}-0.81+\frac{\tilde{n}\tilde{c}}{2}\left[  \ln\left(  l_{s}\tilde
{n}\tilde{c}\right)  -\frac{3}{2}\right]  \right)  \text{ for }\tan\theta
\gg\gamma}{\tilde{n}\ln\frac{\sqrt{1+\left(  \gamma/\tilde{n}\tilde{c}\right)
^{2}}+1}{2}+\frac{\ln\left(  8l_{s}\tilde{n}\tilde{c}\right)  -\gamma_{E}%
}{2\tilde{n}\tilde{c}^{2}}\text{ for }\tan\theta\ll\gamma}%
%EndExpansion
,\label{EnSingle}%
\end{equation}
where $l_{s}=\lambda_{ab}/s$, $\tilde{E}_{PS}^{s}=\tilde{n}\left[  \ln
(\lambda_{ab}/\xi)+0.497\right]  $ is the energy of aligned pancake stacks and
$\tilde{E}_{JV}^{s}=\frac{1}{\tilde{c}}\left(  \ln l_{s}+1.54\right)  $ is the
energy of an isolated Josephson vortices. The first asymptotics corresponds to
the regime of kinked lines. The interaction energy is given by%
\begin{equation}
\tilde{E}_{TV}^{i}\!\approx\!\frac{\pi\gamma}{\tilde{c}^{2}}\!+\!\pi\tilde
{n}^{2}\!-\!\sqrt{\tilde{n}^{2}+\tilde{c}^{-2}}\left[  \!\ln\left(  4\pi
\gamma\sqrt{\tilde{n}^{2}+\tilde{c}^{-2}}\right)  \!-\!\gamma_{E}\right]
+\!\tilde{n}\ln\frac{\gamma\left(  \tilde{n}+\sqrt{\tilde{n}^{2}+\tilde
{c}^{-2}}\right)  }{\tilde{n}+\sqrt{\tilde{n}^{2}+\left(  \gamma/\tilde
{c}\right)  ^{2}}}\label{EnInter}%
\end{equation}
with $\zeta(3)\approx1.202$. This result is valid for $\tan\theta
<c\gamma/(2\pi s)$ covering the relevant angular range.
\end{widetext}

To find equilibrium $\tilde{n}$ for fixed reduced magnetic field strength
$\tilde{h}$, one has to minimize $F_{TV}(\tilde{n})=E_{TV}(\tilde{n}%
)-\tilde{n}\tilde{h}$. In the range of large $\tilde{c}$ above the certain
critical value the function $F_{TV}(\tilde{n})$ has two minima and the
equilibrium pancake density experiences \ a jump between the two values
$\tilde{n}_{\mathrm{dn}}$ and $\tilde{n}_{\mathrm{up}}$  at the critical value
of field . These values are converted to the magnetic inductions using Eq.
(\ref{BxBz}). Figure 6 shows example of the phase diagram obtained assuming
$\lambda_{ab}=200$ nm, $s=1.2$ nm and $\gamma=7$. We also assumed
$\beta=2\sqrt{3}$ in Eq. (\ref{BxBz}).

%\vspace{5mm}
%\maketitle{\bf{Imaging formation of the vortex domains}}
\section{Imaging formation of the vortex domains}

To track the initial stage of the domain nucleation we imaged the flux structure during gradual reduction of the field after initially field cooling the sample to T = 60K with $H_{//}$ = 1388 Oe tilted by $1.6^{o}$ (see Fig. 7).  Before reducing $H_{//}$, the $B_{z}$ distribution (not shown) is homogeneous over the entire imaged area with very faint darker contrast near the edges of the sample due to weak Meissner flux expulsion.  Decreasing field to $H_{//}$ = 832 Oe results in the appearance of a dark edge contrast due to the stray fields coming from the trapped positive $B_{z}$. Simultaneously, stripes of alternating bright and dark contrast (modulated $B_{z}$) emerge along $H_{//}$ near the left and right ends of the sample (Fig. 7a).  Further decreasing the magnetic field, extends the stripes over the entire length of the sample (Fig. 7b). At even smaller $H_{//}$=129 Oe,  vortices turn into the plane near the perimeter of the sample (darker periphery) and a region of trapped $B_{z}$ with stripe domains forms inside the crystal as depicted by the bright contrast in Fig. 7c. This configuration remains after reducing the field to zero (Fig. 7d).  
\vspace{0.5cm}

\textbf{References}
\vspace{0.2cm}

[1s] L.\ L.\ Daemen, L.\ J.\ Campbell, A.\ Yu.\ Simonov, and
V.\ G.\ Kogan Phys.\ Rev.\ Lett.\ \textbf{70}, 2948 (1993); A.\ Sudb{\o },
E.\ H.\ Brandt, and D.\ A.\ Huse Phys.\ Rev.\ Lett.\ \textbf{71}, 1451 (1993);
E.\ Sardella, Physica C \textbf{257}, 231 (1997).

[2s] A. E. Koshelev,  Phys. Rev. B   \textbf{71}, 174507 (2005).

[3s] S. N. Gordeev, A. A. Zhukov, P. A. J. de Groot, A. G.
M. Jansen, R. Gagnon, and L. Taillefer, Phys. Rev. Lett.  \textbf{85}, 4594 (2000).
\begin{widetext}
\begin{center}
\begin{figure}[h]
%\hspace{-12cm}
%\includegraphics{FigA.jpg}
\includegraphics[width=10.0cm]{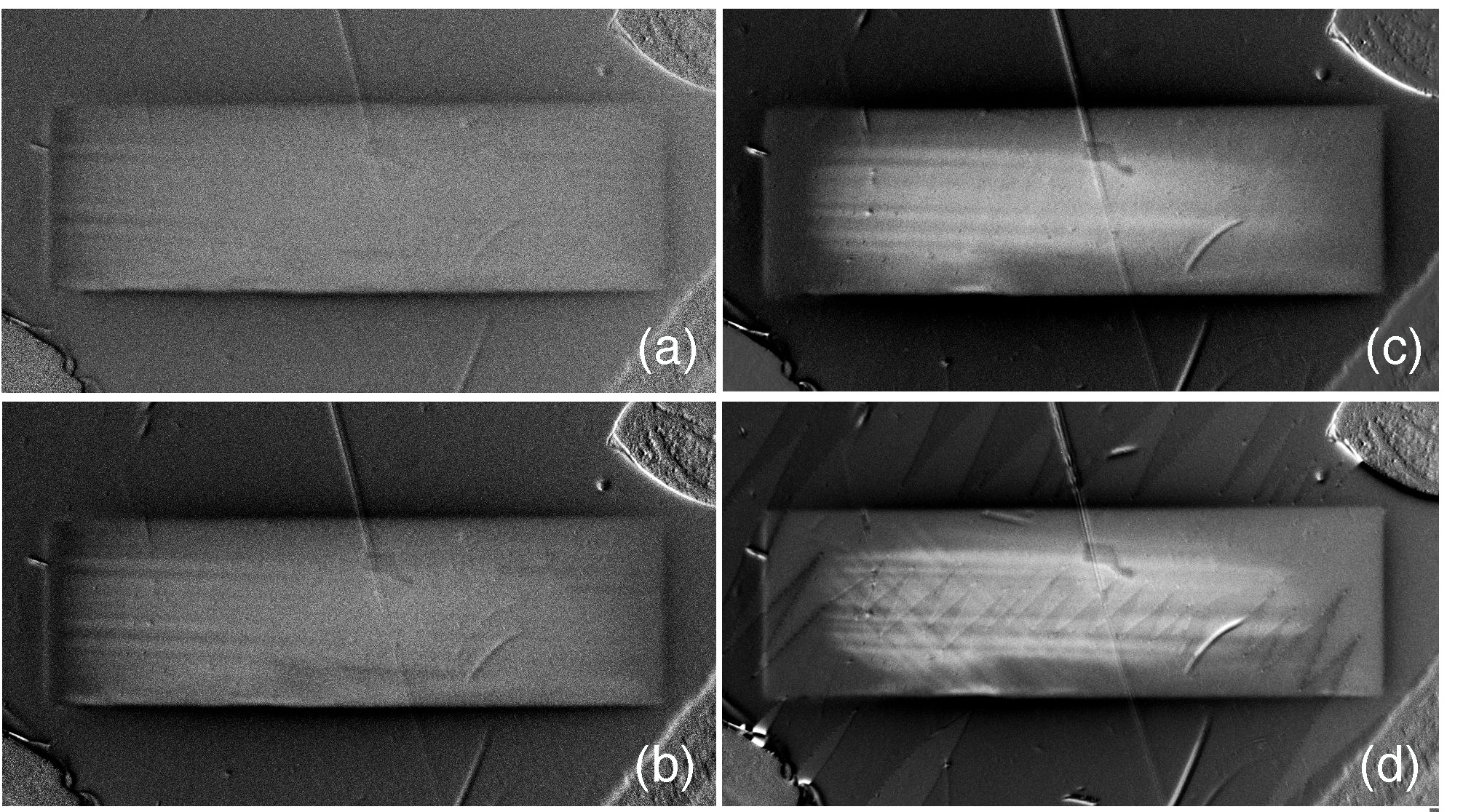}
\caption{Nucleation of domains during reduction of $H_{//}$ at $1.60^{o}$. The sample has been cooled in 1388 Oe to 60K. (a)-(d) $H_{//}$=832, 555, 129, and 0 Oe. The zig-zag features in (d) are magnetic domains in the MO indicator. }
%\label{fig:Fig.A}
\end{figure}
\end{center}
\end{widetext}

\end{document}